\documentclass{svjour3}                     
\smartqed  
\usepackage{graphicx}
\journalname{GERG}
\begin{document}

\title{Continuous matter creation and the acceleration of the universe: a replay}
\titlerunning{replay}        
\author{Alain de Roany  \and J.A. de Freitas Pacheco}

\institute{Universit\'e de Nice-Sophia Antipolis-Observatoire de la C\^ote d'Azur\\
BP 4229, 06304, Nice Cedex 4, France\\
\email{deroany@oca.eu, pacheco@oca.eu}           
}
\date{Received: date / Accepted: date}
\maketitle

\begin{abstract}
In a recent note (arXiv:1012.5069), the investigation performed by the present authors on the evolution 
of density fluctuations in an accelerated universe including matter creation was criticized.
The criticism is based on the fact that the Newtonian background is not "accelerating", invalidating
the conclusions of the linear analysis. We show that our linear equations describe
adequately an accelerating universe in which the pressure associated to the creation process
is constant, a model equivalent to the $\Lambda$CDM cosmology. Thus, our previous
conclusions remain unchanged. 
\end{abstract}

\vspace{2.0cm}

In a recent research note, Lima et al. \cite{lima10} criticize a previous paper by the present 
authors \cite{rp10}, hereafter RP10, who have discussed the possibility
that the observed expansion of the universe is accelerated by a mechanism related to a 
continuous process of matter creation. This is not a new idea, since such a possibility was already 
invoked in the past but within a 
different context, by authors like 
Zeldovich \cite{zel} and  Prigogine \cite{prigo}.

Lima et al. \cite{lima10} compare first both  relativistic solutions derived from Einstein equations, detailed
respectively in references \cite{loj} and \cite{rp10}. It worth mentioning that both approaches, despite the fact that they differ on the 
the adopted {\it ansatz} defining the particle
creation rate, are equivalent since the pressure associated to such a process is constant. In fact, in reference \cite{loj} the 
pressure associated to
the creation process (of ``cold" dark matter)  is given by
\begin{equation} 
\label{pressao1}
P_c = -\frac{\rho\Gamma}{3H}
\end{equation}
where $\rho$ is the energy density, $\Gamma$ is the timescale of particle creation and $H$ is the Hubble 
parameter. In a second step, those authors
assume that the creation rate is proportional to the Hubble parameter, i.e.,
\begin{equation}
\rho\Gamma = 3\alpha\rho_0H
\end{equation}
where $\alpha$ is a constant and $\rho_0$ is the present matter energy density. Clearly, from 
these relations, it results trivially that $P_c = -\alpha\rho_0$ or,
in other words, the thermodynamic pressure associated to matter creation is constant.

In the approach developed by the present authors \cite{rp10}, the thermodynamic formalism developed in \cite{prigo} was
adopted. In this case, they obtained 
for the pressure
associated to the creation process (of ``cold" dark matter)
\begin{equation}
\label{pressao2}
P_c = -\frac{\rho}{3H}\left(3H+\frac{d\lg n}{dt}\right)
\end{equation}
They considered the {\it ansatz} by the which the pressure related to the creation process is 
constant, namely,  $P_c = - \lambda$, then 
one verifies trivially from
eq.~\ref{pressao2} that the particle creation rate is proportional to the Hubble parameter. Thus, in conclusion, the 
world models described either
in reference \cite{loj} or in reference \cite{rp10} are equivalent and behave like the ``standard" $\Lambda$CDM cosmology.
We emphasize this point and the reader should have in mind this aspect when we will discuss 
the criticism made by \cite{lima10} on the linearized equations based on a Newtonian approach.

Nevertheless, it worth mentioning here some important
aspects already emphasized in our original paper \cite{rp10} concerning conceptual differences between 
the ``standard" ($\Lambda$CDM) and the 
continuous creation cosmologies. Firstly, the
solution for the creation pressure expressed by eq.~\ref{pressao2} assumes that the entropy per particle 
remains constant during the expansion and
such a condition {\it is not equivalent to an adiabatic expansion}. In fact, there is a continuous production 
of entropy whose rate is equal to that of particle production, e.g.,
\begin{equation}
\frac{d\lg s}{dt}=\frac{d\lg n}{dt}
\end{equation}
The second point concerns the future evolution of the universe since contrary what will occur in the 
standard model, the matter density doesn't go to zero but 
to a finite value, resulting from the equilibrium between expansion and production rates. Finally, in the 
continuous creation cosmology, the present matter density is
equal to the critical value. This is an important point since it leads to an overestimate of the 
density of cluster of galaxies as it was shown by the present authors \cite{rp10}.

In the second part of their research note, Lima et al. \cite{lima10} discuss the Newtonian 
approach of  \cite{rp10}, concluding that the ``background" 
described by the ``zero"-order equations is not accelerating, thus invalidating the conclusions 
of the linear analysis.

The Newtonian approach is valid only in the weak field limit, useful to describe the evolution of small 
density fluctuations and local velocity
fields but clearly  not adequate to describe the evolution of the scale factor as Lima et al. \cite{lima10} 
tried to demonstrate. 
Certainly, we could have improved our Newtonian approach, introducing "by hand" a modification in the Gauss 
equation, since we know from General Relativity that  pressure acts as a 
source of gravitation. In this case the original eq.~20 in reference \cite{rp10} would have been written as
\begin{equation}
\nabla\cdot\vec{g} = 4\pi G\left(mn+\frac{P}{c^2}\right)
\end{equation}
Since in our particular case $P=P_c=-\lambda$, the equation above can be recast as
\begin{equation}
\label{gauss}
\nabla\cdot\vec{g}=4\pi G\left(mn-\frac{\lambda}{c^2}\right)=4\pi Gmn-\frac{4\pi G\lambda}{c^2}
\end{equation}

Repeating the trivial calculations performed by Lima et al. \cite{lima10}, one obtains now
\begin{equation}
\frac{\ddot a}{a}=-\frac{4\pi G}{3c^2}\left(mn-\lambda\right)
\end{equation}
indicating that an acceleration is possible if $\lambda > mn$. Since the second term on 
the right of eq~\ref{gauss} is constant, the system of linearized
equations derived in reference \cite{rp10} is not modified, describing correctly the evolution of the perturbations 
in an accelerating background. This is not an
 unexpected result, since it is well known from many text books that the cosmological constant does 
not appears explicitly in the linearized equations and the 
present model, with a constant creation pressure is formally equivalent to the $\Lambda$CDM model. In fact,
the main difference between the linear analysis of \cite{rp10} and that of \cite{loj} is that the former 
authors considered  perturbations in the local creation rate 
driven by local variation in the expansion rate, an aspect not taken into account in the 
approach by Lima and collaborators.


\begin{thebibliography}{999}

\bibitem{lima10} J.A.S. Lima, J.F. Jesus and F.A. Oliveira, Note on "Continuous matter
creation and the acceleration of the universe: the growth of density fluctuations, arXiv:1012.5060
\bibitem{rp10} A. de Roany and J.A. de Freitas Pacheco, Continuous matter creation and the acceleration
of the universe: the growth of density fluctuations, Gen.Rel.Grav., online 170D (2010), arXiv:1007.4546
\bibitem{zel} Y. B. Zeldovich, Particle Production in Cosmology, JETP Lett. {\bf 12}, 307 (1970) 	
\bibitem{prigo} I. Prigogine, J. Geheniau, E. Gunzig and P. Nardone,  Thermodynamics and Cosmology, Gen. Rel. 
Grav. {\bf 21}, 767 (1989)
\bibitem{loj} J.A.S. Lima, J.F. Jesus and F.A. Oliveira, CDM accelerating cosmology as an alternative to
$\Lambda$CDM model, JCAP {\bf 1011}, (2010), 027, arXiv:0911.5727
\end{thebibliography}

%

\end{document}